\documentclass[12pt]{article}
\begin{document}

\thispagestyle{empty}
\begin{flushright}
IFIC/04-74, \\ FTUV-05-0101,  \\
1st January 2005
\end{flushright}

\vspace{2cm}

\begin{center}

{\LARGE Generalized curvature and the equations of $D=11$
supergravity}

\vspace{1.5cm}

{\bf  Igor A. Bandos$^{\dagger,\ast}$, Jos\'e A. de
Azc\'arraga$^{\dagger}$,
 Mois\'es Pic\'on$^{\dagger,1}$ and Oscar
Varela$^{\dagger,2}$}

\vspace{1cm} {\it $^{\dagger}$ Departamento de F\'{\i}sica
Te\'orica, Univ.~de Valencia and IFIC (CSIC-UVEG), 46100-Burjassot
(Valencia), Spain
\\
$^{\ast}$Institute for Theoretical Physics, NSC ``Kharkov
Institute of Physics  and Technology'',  UA61108, Kharkov, Ukraine
\\
$^{1}$Department of Physics and Astronomy, University of Southern
California, Los Angeles, CA 90089-2535, USA
\\
$^{2}$Michigan Center for Theoretical Physics, Randall Laboratory,
Department of Physics, University of Michigan, Ann Arbor, MI
48109-1120, USA }
\\
\vskip 4cm

\end{center}

\abstract{It is known that, for zero fermionic sector,
$\psi_\mu^{\alpha}(x)=0$, the bosonic equations of
Cremmer--Julia--Scherk eleven--dimensional supergravity can be
collected in a compact expression, ${\cal
R}_{ab}{}_{\alpha}{}^{\gamma}\Gamma^b{}_{\gamma}{}^{\beta}=0$,
which is a condition   on the curvature ${\cal
R}_{\alpha}{}^{\beta}$ of the generalized connection $w$. In this
letter we show that the equation ${\cal
R}_{bc}{}_{\alpha}{}^{\gamma} \Gamma^{abc}{}_{\gamma\beta}= 4i
((\hat{{\cal D}}\psi)_{bc}\Gamma^{[abc})_\beta \,
(\psi_d\Gamma^{d]})_\alpha$, where $\hat{{\cal D}}$ is the
covariant derivative for  the generalized connection $w$, collects
all the bosonic equations of $D=11$ supergravity when the
gravitino is nonvanishing, $\psi_\mu^{\alpha}(x)\not=0$. }

\newpage

\section{Introduction}

Recently, the notion of generalized connection and generalized
holonomy has been applied to the analysis of supersymmetric
solutions of $D=10,11$ dimensional supergravity
\cite{Duff03,Hull03,FFP02,GP02,P+T03,BDLW03,BPS03}. The
generalized connection (see  \cite{CJS})
\begin{eqnarray}
\label{w=om+t} w_\beta{}^\alpha := \omega_{L \beta}{}^\alpha +
t_1{}_\beta{}^\alpha = {1\over 4}  \omega^{ab}
\Gamma_{ab\,\beta}{}^\alpha + t_1{}_\beta{}^\alpha
\end{eqnarray}
involves, in addition to the true Lorentz (or spin) connection,
the Lorentz covariant part
\begin{eqnarray}
\label{t=gh}  t_1{}_\beta{}^\alpha &=&   {i\over 18} E^a
\left({{}\over {}} F_{a[3]} \Gamma^{[3]}{}_\beta{}^\alpha +
{1\over 8}F^{[4]} \Gamma_{a[4]}{}_\beta{}^\alpha \right) \; ,
\end{eqnarray}
constructed from the tensor $F_{abcd}$, the `supersymmetric' field
strength of the antisymmetric tensor field $A_{\mu\nu\rho}(x)$
(see Eqs. (\ref{dA3=a+F})). In Eq. (\ref{t=gh})  $ F_{a[3]}
\Gamma^{[3]} :=F_{ab_1b_2b_3} \Gamma^{b_1b_2b_3}$, $ F^{[4]}
\Gamma_{a[4]}{}_{\beta}{}^{\alpha} = F^{b_1\ldots b_4}
\Gamma_{ab_1\ldots b_4}{}_{\beta}{}^{\alpha}$ and we have denoted
the vielbein one--form $dx^\mu e_\mu^a(x)$ by $E^a$, $E^a = dx^\mu
e_\mu^a(x)$.

The generalized connection allows for a simple expression of the
supersymmetric  transformation rules for the gravitino (hence the
name of `supersymmetric' connection frequently  used). Denoting
the gravitino one--form by $\psi^\alpha= dx^\mu
\psi_\mu^{\alpha}(x)$, this variation is given by
\begin{eqnarray}
\label{susypsi=} \delta_{\varepsilon}\psi^\alpha = {\cal
D}\varepsilon^\alpha(x)&:=& D\varepsilon^\alpha(x)
-\varepsilon^\beta(x) t_{1 \beta}{}^\alpha(x) = \nonumber \\
&=& d\varepsilon^\alpha(x) - \varepsilon^\beta(x)
w_\beta{}^\alpha(x) \; . \qquad
\end{eqnarray}

It was already noticed in \cite{CJS} that the gravitino equation
of motion has also a compact form (see  Eq. (\ref{Eqmpsi})) in
terms of its generalized covariant (or `supercovariant')
derivative
\begin{eqnarray}\label{hDpsi}
\hat{{\cal D}} \psi^\alpha &:=& d\psi^\alpha - \psi^\beta \wedge
w_\beta{}^\alpha \equiv  D\psi^\alpha - \psi^\beta \wedge
t_1{}_\beta{}^\alpha \;
\end{eqnarray}
 defined for  the generalized connection (\ref{w=om+t}).
Then the following observation (see \cite{GP02,BPS03}) holds: when
the fermionic sector is set to zero, {\it all} the {\it bosonic}
equations of the Cremmer--Julia--Scherk (CJS) eleven--dimensional
supergravity can be collected in the  simple expression
\begin{eqnarray}\label{bEqm.}
{\cal N}_{a\, \beta}{}^{\alpha} := {\cal
R}_{ab}{}_\beta{}^{\gamma} \Gamma^b{}_\gamma{}^{\alpha} =0 \;
\end{eqnarray}
or, equivalently, $i_b {\cal R}_\beta{}^{\gamma}
\Gamma^b{}_\gamma{}^{\alpha}\equiv E^a{\cal
R}_{ba}{}_\beta{}^{\gamma} \Gamma^b{}_\gamma{}^{\alpha} =0$,  in
terms of the generalized curvature ${\cal R}$ (see, e.g.
\cite{FFP02})
\begin{eqnarray}\label{calR}
{\cal R}_\beta{}^{\alpha} &:=& dw_\beta{}^{\alpha} - w
_\beta{}^{\gamma}\wedge  w_\gamma{}^{\alpha} \nonumber \\
&=& {1\over 4} R^{ab}(\Gamma_{ab})_{\alpha}{}^{\beta} +
Dt_1{}_{\alpha}{}^{\beta} - t_{1 \, \alpha}{}^\gamma \wedge t_{1\,
\gamma}{}^{\beta} \; , \qquad
\end{eqnarray}
which takes values in the Lie algebra of the generalized holonomy
(holonomy of the generalized connection) group \cite{Duff03}
\footnote{See \cite{Oscar+} for further discussion on the
generalized holonomy.}. A similarly concise equation in the case
of the purely bosonic limit of massive type IIA supergravity was
given recently in \cite{LustTsimpis}.

We present here the generalization of Eq. (\ref{bEqm.}), to the
case of nonzero gravitino,  $\psi^\alpha\not=0$. It reads
\begin{eqnarray}
\label{SfEq0.} {\cal R}_\beta{}^\gamma \wedge E^{\wedge 8}_{abc}
\Gamma^{abc}_{\gamma\alpha} = -i \hat{{\cal D}} \psi^\delta \wedge
\psi^\gamma \wedge  E^{\wedge 7}_{a_1\ldots a_4}
\Gamma^{[a_1a_2a_3}_{\delta\alpha} \Gamma^{a_4]}_{\beta\gamma} \;
 \;\; \\ \label{SfEq1.}  \Rightarrow \; \;   {\cal
R}_{bc}{}_{\alpha}{}^{\gamma} \Gamma^{abc}{}_{\gamma\beta} = 4i
((\hat{{\cal D}}\psi)_{bc}\Gamma^{[abc})_\beta \,
(\psi_d\Gamma^{d]})_\alpha \; , \quad
\end{eqnarray}
where $\hat{{\cal D}} \psi^\alpha = 1/2 E^a \wedge E^b (\hat{{\cal
D}}\psi)_{ba}{}^\alpha$ is defined in (\ref{hDpsi}) and
\begin{eqnarray}
\label{E11-n} E^{\wedge (11-k)}_{a_1\ldots a_k} &
:=\frac{1}{(11-k)!} \varepsilon_{a_1\ldots a_kb_1\ldots b_{11-k}}
E^{b_1}\wedge \ldots \wedge E^{b_{11-k}} \; .
\end{eqnarray}
 Eq. (\ref{SfEq0.}) (or  (\ref{SfEq1.}))
 collects all the bosonic equations when the gravitino  is not zero,
 $\psi^\alpha\not=0$.

Although the final result is formulated as a statement about
dynamical equations of motion and, in this sense, refers to the
second order approach to supergravity, we find it convenient to
 use the first order supergravity action of \cite{D'A+F,J+S99}.
 Our notation (which is
explained in the text) is close to that in \cite{J+S99} and the
same of \cite{BPS03,AnnP04}.

\section{First order action for $D=11$ supergravity }

The first order action for $D=11$ supergravity \cite{D'A+F,J+S99},
\begin{eqnarray}\label{S11=}
S=\int_{{\cal M}^{11}}{\cal L}_{11}[E^a, \psi^\alpha, \omega^{ab},
A_3 , F_{a_1a_2a_3a_4}] \; ,
\end{eqnarray}
is the integral over eleven--dimensional spacetime $M^{11}$ of the
eleven form ${\cal L}_{11}$ which can be written as
\cite{D'A+F,J+S99}
\begin{eqnarray}\label{L11=}
{\cal L}_{11} &=& {1\over 4} R^{ab}\wedge E^{\wedge 9}_{ab} -
D\psi^\alpha \wedge \psi^\beta   \wedge
\bar{\Gamma}^{(8)}_{\alpha\beta} + {1\over 4}  \psi^\alpha \wedge
\psi^\beta   \wedge (T^a + i/2 \, \psi \wedge \psi \, \Gamma^a)
\wedge E_a
\wedge \bar{\Gamma}^{(6)}_{\alpha\beta}  + \nonumber \\
&+& (dA_3- a_4) \wedge (\ast F_4 + b_7) +  {1\over 2} a_4 \wedge
b_7  -
 {1\over 2} F_4 \wedge \ast F_4  -
{1\over 3} A_3 \wedge dA_3\wedge dA_3 \; .
\end{eqnarray}
Following \cite{J+S99} (see also \cite{AnnP04}), we have
introduced the notation
\begin{eqnarray}\label{a4}
a_4&:= {1\over 2} \psi^\alpha \wedge \psi^\beta   \wedge
\bar{\Gamma}^{(2)}_{\alpha\beta}\; ,  \quad   b_7:= {i\over 2}
\psi^\alpha \wedge \psi^\beta   \wedge
\bar{\Gamma}^{(5)}_{\alpha\beta} \;  \quad
\end{eqnarray}
for the bifermionic 4-- and 7--forms and
\begin{eqnarray}\label{F4:=}
& F_4 := {1\over 4!} E^{a_4} \wedge \ldots \wedge E^{a_1}
F_{a_1\ldots a_4} \; , \quad \nonumber \\ & \ast F_4:= - {1\over
4!} E^{\wedge 7}_{a_1\ldots a_4} F^{a_1\ldots a_4} \; . \qquad
\end{eqnarray}
for the purely bosonic forms constructed from the antisymmetric
tensor zero-form $F_{abcd}$.
 We also use the compact notation
\begin{eqnarray}
\label{Gammak} \bar{\Gamma}^{(k)}_{\alpha\beta} &:=& {1\over k!}
E^{a_k} \wedge \ldots \wedge E^{a_1} \Gamma_{a_1 \ldots a_k}
{}_{\alpha\beta} \;
\end{eqnarray}
 and Eq. (\ref{E11-n}) [to be compared with  the notation of \cite{J+S99},
 $E^{\wedge (11-k)}_{a_1\ldots
a_k} = \Sigma_{a_1\ldots a_k}$, $\bar{\Gamma}^{(k)}_{\alpha\beta}
= (-)^{k(k-1)/2} \gamma^{(k)}_{\alpha\beta}$].

The action (\ref{S11=}) is invariant under the local supersymmetry
 transformations  $\delta_{\varepsilon}$, which are given by
\begin{eqnarray}
\label{susye} \delta_{\varepsilon}E^a &=& - 2i  {\psi}^\alpha
\Gamma^a_{\alpha\beta}{\varepsilon}^\beta \; ,
\\
\label{susyf} \delta_{\varepsilon}\psi^\alpha &=& {\cal
D}\varepsilon^\alpha(x) = D\varepsilon^\alpha(x) -
\varepsilon^\beta(x) t_{1\beta}{}^\alpha(x) \; ,  \qquad
\\
\label{susyA} \delta_{\varepsilon}A_3 &=& \psi^\alpha \wedge
\bar{\Gamma}^{(2)}_{\alpha\beta}{\varepsilon}^\beta \; ,
\end{eqnarray}
plus more complicated expressions for
$\delta_{\varepsilon}\omega^{ab}$ and
$\delta_{\varepsilon}F_{abcd}$,  which can be found in
\cite{J+S99} and that will not be needed below.  Let us stress
that, as shown in \cite{J+S99}, the supersymmetry transformation
rules of the physical fields are the same in the second and in the
first order formalisms.

\section{Equations of motion}

In the first order action (\ref{S11=}) one distinguishes between
the true equations of motion and the  {\it algebraic} (or {\it
nondynamical}) equations
\begin{eqnarray}
\label{STa=} T^a & = & -i \psi^\alpha \wedge \psi^\beta \,
\Gamma^a_{\alpha\beta} \; ,
\\ \label{dA3=a+F}
dA_3 &=&  a_4 + F_4 \;  \quad
\end{eqnarray}
(see Eqs. (\ref{a4}), (\ref{F4:=}) for the notation) which follow,
respectively, from the variation with respect to the spin
connection $\omega_L^{ab}$  and the antisymmetric tensor
$F_{abcd}$
\begin{eqnarray}
\label{vomL} \delta_{\omega} {\cal L}_{11} = {1\over 4} E^{\wedge
8}_{abc} \wedge (T^a + i\psi^{\alpha}\wedge \psi^{\beta}\,
\Gamma^a_{\alpha\beta}) \wedge \delta \omega^{bc} + d(\ldots) \; ,
\nonumber \\
  \delta_{F} {\cal L}_{11} =  - {1\over 4!}
 (dA_3- a_4-F_4) \wedge
E^{\wedge 7}_{a_1\ldots a_4} \, \delta  F^{a_1\ldots a_4}\, .
\qquad
\end{eqnarray}
 Notice that Eqs.
(\ref{STa=}), (\ref{dA3=a+F}) are the counterparts of the
superspace constraints of $D=11$ supergravity (see \cite{AnnP04}
for a  discussion and references), but for  forms on
eleven--dimensional spacetime. As far as  the {\it dynamical
bosonic equations} are concerned, one finds that the variation
with respect to $A_3$,
\begin{eqnarray}
\label{vA3L} \delta_{A} {\cal L}_{11} &=& d(\ast F_4 + b_7
-A_3\wedge dA_3) \wedge \delta A_3 + d(\ldots)  \qquad
\end{eqnarray}
results in
\begin{eqnarray}
\label{EqmA3} {\cal G}_8 &:=&  d(\ast F_4 + b_7 -A_3\wedge dA_3)
=0 \; ,
\end{eqnarray}
which becomes the standard CJS
 three--form gauge field equations of motion once the
 algebraic equations (constraints) (\ref{STa=}), (\ref{dA3=a+F}) are taken into
 account.
The more complicated variation $\delta_{E}$ with respect to the
vielbein form, $ \delta_{E} {\cal L}_{11}= \ldots$, as well as the
full expression  of the Einstein equations
\begin{eqnarray}
\label{EqmE} M_{10\, a}  &:=& {1\over 4} R^{bc} \wedge E^{\wedge
8}_{abc} + \ldots \; = 0 \; ,
\end{eqnarray}
which follows from that variation, will not be needed here (see
\cite{J+S99}).

After some algebra,  the  fermionic variation $\delta_{\psi}$ of
the Lagrangian form ${\cal L}_{11}$, Eq. (\ref{L11=}), reads ({\it
cf.} \cite{J+S99})
\begin{eqnarray}
\label{vpsiL} \delta_{\psi} {\cal L}_{11} = -2 \hat{{\cal D}}
\psi^\alpha \wedge \bar{\Gamma}^{(8)}_{\alpha\beta}  \wedge \delta
\psi^\beta
 + i (dA_3- a_4-F_4) \wedge
\bar{\Gamma}^{(5)}_{\alpha\beta}  \wedge \psi^\alpha \wedge \delta
\psi^\beta + \nonumber \\ +  \left( i_a
\bar{\Gamma}^{(8)}_{\alpha\beta}  + 1/2 E_a \wedge
\bar{\Gamma}^{(6)}_{\alpha\beta} \right) \wedge (T^a +
i\psi^{\alpha}\wedge \psi^{\beta}\, \Gamma^a_{\alpha\beta}) \wedge
\psi^\alpha \wedge  \delta \psi^\beta -  \nonumber \\
 - d\; \left[ \psi^\alpha \wedge \bar{\Gamma}^{(8)}_{\alpha\beta}
\wedge \delta \psi^\beta \right] \; ,
\end{eqnarray}
where $\hat{{\cal D}} \psi^\alpha$ is given by Eq. (\ref{hDpsi}).
Taking  into account the algebraic equations (\ref{STa=}),
(\ref{dA3=a+F}), and ignoring the (last) total derivative term in
Eq. (\ref{vpsiL}) one finds the {\it gravitino equation} of
\cite{CJS}  written, as in \cite{J+S99}, in
 the  suggestive differential form
\begin{eqnarray}
 \label{Eqmpsi} \Psi_{10\; \beta}  &:=& \hat{{\cal D}}
\psi^\alpha \wedge \bar{\Gamma}^{(8)}_{\alpha\beta}  =0 \; .
\end{eqnarray}

\section{Bosonic equations of $D=11$ supergravity as a condition
on the generalized curvature}

\subsection{A concise  form of the bosonic
equations from selfconsistency of the gravitino equations}

It is important that the above gravitino equation, $\Psi_{10\;
\beta}=0$,   is expressed in terms of the covariant derivative
$\hat{{\cal D}}$, Eqs. (\ref{hDpsi}), (\ref{w=om+t}),
(\ref{t=gh}). As a result,  the integrability/selfconsistency
condition for Eq.  (\ref{Eqmpsi}) may be written in terms of the
generalized curvature (\ref{calR}). Using $\hat{{\cal
D}}\hat{{\cal D}} \psi^\alpha = - \psi^\beta \wedge {{\cal
R}}_\beta{}^\alpha$ and $t^{\quad\gamma}_{1[\beta} \wedge
\bar{\Gamma}^{(8)}{}_{\alpha]\gamma} = 0$  \footnote{This follows
{\it e.g.},  from direct calculation of $ t^{\quad\gamma}_{1\,
\alpha} \wedge \bar{\Gamma}^{(8)}_{\gamma\beta} = -{i\over 2} F_4
\wedge \bar{\Gamma}^{(5)}_{\alpha\beta}+ {1\over 2} \ast F_4
\wedge \bar{\Gamma}^{(2)}_{\alpha\beta}$.} which implies ${\cal D}
\bar{\Gamma}^{(8)}_{\beta\alpha} = {D}
\bar{\Gamma}^{(8)}_{\beta\alpha} = T^a \wedge i_a
\bar{\Gamma}^{(8)}_{\beta\alpha}$, we obtain
\begin{eqnarray}
\label{DEqmpsi} \hat{{\cal D}} \Psi_{10\; \alpha} &= & \hat{{\cal
D}} \psi^\beta \wedge (T^a + i \psi \wedge \psi \Gamma^a) \wedge
i_a \bar{\Gamma}^{(8)}_{\beta\alpha} - \nonumber \\ && - {i \over
6} \psi^\beta \wedge \left[ {\cal R}_\beta{}^\gamma \wedge
E^{\wedge 8}_{abc} \Gamma^{abc}_{\gamma\alpha}
 + i \hat{{\cal D}} \psi^\delta \wedge \psi^\gamma \wedge
E^{\wedge 7}_{a_1\ldots a_4} \Gamma^{[a_1a_2a_3}_{\delta\alpha}
\Gamma^{a_4]}_{\beta\gamma}\right] = 0\; . \qquad
\end{eqnarray}
The first term in the second part of Eq. (\ref{DEqmpsi}) vanishes
due to the algebraic (constraint) equation  (\ref{STa=}). Hence on
the surface of constraints the selfconsistency of the gravitino
equation is guaranteed when
\begin{eqnarray}
\label{SfEq} {\cal M}_{10\; \alpha\beta} := {\cal
R}_\beta{}^\gamma \wedge E^{\wedge 8}_{abc}
\Gamma^{abc}_{\gamma\alpha} + i \hat{{\cal D}} \psi^\delta \wedge
\psi^\gamma \wedge  E^{\wedge 7}_{a_1\ldots a_4}
 \Gamma^{[a_1a_2a_3}_{\delta\alpha} \Gamma^{a_4]}_{\beta\gamma} = 0 \; . \qquad
\end{eqnarray}
Our main observation  is that
 {\it Eq. (\ref{SfEq}) (see Eq. (\ref{SfEq0.}) or (\ref{SfEq1.}))  collects all the bosonic
equations of motion (\ref{EqmA3}), (\ref{EqmE}) and the
corresponding Bianchi identities for the $A_3$ gauge field and for
the Riemann curvature tensor.} Let us stress that we distinguish
between the algebraic equations or constraints, Eqs. (\ref{STa=})
and (\ref{dA3=a+F}), from the true dynamical equations
(\ref{EqmA3}), (\ref{EqmE}),  and that our statement above refers
to the dynamical equations; thus it is also true for the second
order formalism.

 To show this it is not necessary to make
an explicit calculation. It is sufficient to use the second
Noether theorem and/or the fact that the purely bosonic limit of
(\ref{SfEq}) implies Eq. (\ref{bEqm.}) (see Sec. 4.3 below), which
is equivalent to the set of all bosonic equations and Bianchi
identities when $\psi^\alpha=0$.

\subsection{Proof using
the Noether identities for supersymmetry}

In accordance with the second Noether theorem, the local
supersymmetry under (\ref{susye})--(\ref{susyA}) reflects (and is
reflected by) the existence of an interdependence among the
bosonic and fermionic equations of motion; such a relation is
called a Noether identity. Furthermore, as the local supersymmetry
variation of the gravitino is given by the covariant derivative
$\hat{{\cal D}}\varepsilon^\alpha$ with generalized connection,
Eq. (\ref{susyf}), the gravitino equation $\Psi$ should enter the
corresponding Noether identity through $\hat{{\cal D}}\Psi$. Thus,
$\hat{{\cal D}}\Psi$ should be expressed in terms of the equations
of motion for the bosonic fields, in our case including the
algebraic equations for the auxiliary fields. Hence, in the light
of (\ref{DEqmpsi}), (\ref{STa=}) the {\it l.h.s.} of Eq.
(\ref{SfEq}) vanishes when {\it all} the bosonic equations are
taken into account.

Indeed, schematically, ignoring for simplicity the purely
algebraic equations and neglecting the boundary contributions, the
variation of the action (\ref{S11=}), (\ref{L11=}) (considered now
in the second order formalism) reads
\begin{eqnarray}
\label{vS11=} \delta S = \int \left(-2\Psi_{10\, \alpha} \wedge
\delta  \psi^\alpha + {\cal G}_8 \wedge \delta A_3 + M_{10 \, a}
\wedge \delta E^a \right) \; . \qquad
\end{eqnarray}
For the local supersymmetry transformations
$\delta_{\varepsilon}$, Eqs. (\ref{susyf}), (\ref{susye}) and
(\ref{susyA}), one finds
 integrating by parts
\begin{eqnarray}
\label{vsusyS11=0} \delta_{\varepsilon} S &=& \! \int
\!\left(-2\Psi_{10\, \alpha} \wedge {\cal D} {\varepsilon}^\alpha
+ {\cal G}_8 \wedge \delta_{\varepsilon}A_3 + M_{10 \, a} \wedge
\delta_{\varepsilon} E^a \right) =  \nonumber \\ && = - \int
(-2{\cal D}\Psi_{10\, \alpha} -  {\cal G}_8 \wedge
 \psi^\beta \wedge
\bar{\Gamma}^{(2)}_{\beta\alpha}
  +2i  M_{10 \, a} \wedge
 \psi^\beta  \Gamma^{a}_{\beta\alpha}
) \, \varepsilon^\alpha  =0 \, . \qquad
\end{eqnarray}
As  $\delta_{\varepsilon} S=0$ is satisfied for an arbitrary
fermionic function $\varepsilon^\alpha(x)$, it follows that
\begin{eqnarray}
\label{DPsi=1} {\cal D}\Psi_{10\, \alpha} = -\frac{1}{2}
\psi^\beta \wedge \left( - 2i \Gamma^{a}_{\beta\alpha} M_{10 \, a}
+  {\cal G}_8 \wedge \bar{\Gamma}^{(2)}_{\beta\alpha} \right) \; .
\quad
\end{eqnarray}

In the light of Eqs. (\ref{DEqmpsi}) and (\ref{DPsi=1}), and after
the algebraic equations (\ref{STa=}), (\ref{dA3=a+F})  are taken
into account,
\begin{eqnarray}
\label{SfEq=} {\cal M}_{10\; \alpha\beta} &:=&   {\cal
R}_\beta{}^\gamma \wedge E^{\wedge 8}_{abc}
\Gamma^{abc}_{\gamma\alpha} + i \hat{{\cal D}} \psi^\delta \wedge
\psi^\gamma \wedge  E^{\wedge 7}_{a_1\ldots a_4}
 \Gamma^{[a_1a_2a_3}_{\delta\alpha} \Gamma^{a_4]}_{\beta\gamma} =   \nonumber \\
 & & =-3i \left( - 2i \Gamma^{a}_{\beta\alpha}
M_{10 \, a} +  {\cal G}_8 \wedge \bar{\Gamma}^{(2)}_{\beta\alpha}
\right) \; . \qquad
\end{eqnarray}
It then follows that ${\cal M}_{10\; \alpha\beta}=0$,  Eq.
(\ref{SfEq}), is satisfied after the dynamical equations
(\ref{EqmE}), (\ref{EqmA3}) are used. Moreover, Eq. (\ref{SfEq=})
also shows what Lorentz--irreducible parts of the concise bosonic
equations ${\cal M}_{10\; \alpha\beta}=0$ coincide with the
Einstein and with the 3--form gauge field equations. These are
given, respectively,  by
\begin{eqnarray}
\label{SfEq=Ei}
 && M_{10 \, a}= -{1\over 192} \mathrm{tr} (\Gamma_a{\cal M}_{10}) \; ,
 \qquad \\ \label{SfEq=GF} && {\cal G}_8 \wedge E^a \wedge E^b
 = {i\over 96} \mathrm{tr} (\Gamma^{ab} {\cal M}_{10}) \;  . \qquad
\end{eqnarray}
It is clear that all other Lorentz--irreducible parts in Eq.
(\ref{SfEq}), ${\cal M}_{10\; \alpha\beta}=0$, are satisfied
either identically or due to the Bianchi identities that are the
integrability conditions for the algebraic equations
 (\ref{STa=}), (\ref{dA3=a+F})  used in the derivation of
(\ref{SfEq=}).

Thus, we have proven  that  Eq. (\ref{SfEq}) collects all the
dynamical bosonic equations of motion in the second order approach
to supergravity. To see that it collects all the Bianchi
identities as well, one may either perform a straightforward
calculation or study the pure bosonic limit of Eq. (\ref{SfEq}).
The latter way is simpler and it also provides an alternative
proof of the above statement as we now show below.

 \subsection{Proof
using the purely bosonic limit of the equations}

For bosonic configurations,
$\psi^\alpha=0$, Eq. (\ref{SfEq}) takes the form
\begin{eqnarray}
\label{SfEq0} \psi^\alpha=0 \; , \qquad {\cal R}_\beta{}^\gamma
\wedge E^{\wedge 8}_{abc} \Gamma^{abc}_{\gamma\alpha} =  0 \; .
\end{eqnarray}
We show here that this equation is another form of  Eq.
(\ref{bEqm.})  \cite{BPS03},
\begin{eqnarray}
\label{SfEq01} \psi^\alpha=0 \; , \qquad i_a {\cal
R}_\beta{}^\gamma \Gamma^{a}_{\gamma}{}^{\alpha} \equiv E^b {\cal
R}_{ab\beta}{}^\gamma \Gamma^{a}_{\gamma}{}^{\alpha} =  0 \; .
\end{eqnarray}
As  the above equation (\ref{SfEq01}) collects all the bosonic
equations of standard CJS supergravity as well as all the Bianchi
identities in the purely bosonic limit \cite{GP02,BPS03}, the
equivalence of Eqs. (\ref{SfEq01}) and (\ref{SfEq0})
 will imply that
${\cal M}_{10\; \alpha\beta}=0$, Eq. (\ref{SfEq}), does the same
for the case of nonvanishing fermions, $\psi^\alpha\not=0$.

Decomposing ${\cal R}_\alpha{}^\beta$ on the basis of bosonic
vielbeins, ${\cal R}_\alpha{}^\beta= 1/2 E^a \wedge E^b {\cal
R}_{ba\; \alpha}{}^\beta$, one finds that Eq. (\ref{SfEq0})
implies
\begin{eqnarray}
\label{EqcRb1} {\cal R}_{ab\; \beta}{}^\gamma
\Gamma^{abc}_{\gamma\alpha} =0 \; .
\end{eqnarray}
Contracting (\ref{EqcRb1}) with $\Gamma_{c}^{\alpha\delta}$ one
finds
\begin{eqnarray}
\label{EqcRb2} {\cal R}_{ab\; \beta}{}^\gamma
\Gamma^{ab}{}_{\gamma}{}^{\delta} =0 \; .
\end{eqnarray}
Then, contracting again with the Dirac matrix
$\Gamma_{d}^{\alpha\delta}$ and using $\Gamma^{ab}\Gamma_{d}\Gamma^{ab}{}_{d} + 2 \Gamma^{[a}\delta_d{}^{b]}$  as well as Eq.
(\ref{EqcRb1}), one recovers Eq. (\ref{bEqm.}), ${\cal
N}_{a\beta}{}^\alpha=0$, which is an equivalent form of Eq.
(\ref{SfEq01}).

The Bianchi identities $R_{[ab\; c]d}\equiv 0$ and $dF_4\equiv 0$
appear as the irreducible parts $tr(\Gamma_{c_1c_2c_3}{\cal N}_a)$
and  $tr(\Gamma_{c_1\ldots c_5}{\cal N}_a)$ of Eq. (\ref{bEqm.})
\cite{BPS03} [more precisely, in the last case the relevant part
in  ${\cal N}_a$ is proportional to $[dF_4]_{b_1\ldots b_5}
(\Gamma_a{}^{b_1\ldots b_5} + 10 \delta_a{}^{[b_1}
\Gamma^{b_2\ldots b_5]})$, but the two terms in the brackets are
independent]. Knowing this, one may also reproduce the  terms that
include the Bianchi identities in the concise equation
(\ref{SfEq0.}) (equivalent to (\ref{SfEq1.}) or (\ref{SfEq=}))
with a nonvanishing gravitino.

\section{Conclusion}

We have shown that {\it all the bosonic equations of $D$=11
supergravity can be collected in a single equation, Eq.
(\ref{SfEq})}, written in terms of the generalized curvature
(\ref{calR}) which takes values in the algebra of the generalized
holonomy group. In the first proof we used the shortcut provided
by the second Noether theorem, which implies the Noether identity
(\ref{DPsi=1}) for the local supersymmetry
(\ref{susye})--(\ref{susyA})  relating the bosonic and fermionic
equations. The second proof uses the purely bosonic limit
(\ref{SfEq01}) of the desired equation (\ref{SfEq}). This is
simpler since  the properties of the purely bosonic Eq.
(\ref{SfEq01}) are known \cite{GP02,BPS03} and may also be used to
simplify the extraction of Bianchi identities from Eq.
(\ref{SfEq}), although we do not do it here.

The concise form (\ref{SfEq}) of all the bosonic equations is
obtained by factoring out  the fermionic one--form $\psi^\beta$ in
the  selfconsistency (or integrability) conditions ${\cal
D}\Psi_{10\; \beta}=0$ [Eqs. (\ref{DEqmpsi})], for  the gravitino
equations $\Psi_{10\, \alpha}=0$, Eqs. (\ref{Eqmpsi}). In this
sense, one can say that in (the second order formalism of) {\it
$D=11$ CJS supergravity all the equations of motion and Bianchi
identities are encoded in the fermionic gravitino equation}
 $\Psi_{10\; \beta}  := \hat{{\cal D}}
\psi^\alpha \wedge \bar{\Gamma}^{(8)}_{\alpha\beta} =0$ [Eq.
(\ref{Eqmpsi})].

Actually  this should be expected for a supergravity theory
including  only one fermionic field, the gravitino,  and whose
supersymmetry algebra closes on shell.  As we have
 discussed, the  basis for such an expectation
is provided by the second Noether theorem.

We hope that the explicit form
 (\ref{SfEq}) of the equation collecting all the bosonic equations of motion
 and Bianchi identities may  be useful in a further understanding of the
 properties of $D=11$ supergravity and in the
analysis of its supersymmetric solutions, including those  with
nonvanishing fermionic sector (see \cite{Hull84} and refs
therein).

\bigskip

{\it Acknowledgments}. The authors thank Dima Sorokin for useful
comments and Paul de Medeiros for correspondence. This work has
been partially supported by the research grant BFM2002-03681 from
the Ministerio de Educaci\'on y Ciencia and from EU FEDER funds,
the Generalitat Valenciana (Grupos 03/124), the grant N 383 of the
Ukrainian State Fund for Fundamental Research, the INTAS Research
Project N 2000-254 and the EU network MRTN--CT--2004--005104
`Forces Universe'.
 M.P. and O.V. wish to thank the Ministerio de Educaci\'on y Ciencia and
the Generalitat Valenciana, respectively, for their FPU and FPI
research grants,  and I. Bars (M.P.) and M. Duff (O.V.) for their
hospitality at the USC and the University of Michigan.

\end{document}